\documentclass[preprint,12pt]{elsarticle}

\usepackage{amssymb}
\usepackage{graphicx}
\usepackage{graphics}
\usepackage{rotating}
\usepackage{longtable}
\usepackage{lscape}
\usepackage{epsfig}
\usepackage{rotating}
\usepackage{ifthen}
\usepackage{multicol}
\usepackage{amssymb}
\usepackage{xspace}

\begin{document}

\begin{frontmatter}


\title{Xenon purity analysis for EXO-200 via mass spectrometry}

\author[umd]{A.~Dobi}
\author[umd]{C.~Hall}
\author[umd]{S.~Slutsky}
\author[umd]{Y.-R.~Yen}
\author[laurentian]{B.~Aharmin}
\author[bern]{M.~Auger}
\author[stanford]{P.S.~Barbeau} 
\author[csu]{C.~Benitez-Medina}
\author[slac]{M.~Breidenbach}
\author[laurentian]{B.~Cleveland}
\author[slac]{R.~Conley} 
\author[umass]{J.~Cook} 
\author[csu]{S.~Cook}
\author[stanford]{I.~Counts}
\author[slac]{W.~Craddock} 
\author[umass]{T.~Daniels}
\author[umd]{C.G.~Davis}
\author[stanford]{J.~Davis}
\author[stanford]{R.~deVoe}
\author[carleton]{M.~Dixit}  
\author[stanford]{M.J.~Dolinski} 
\author[laurentian]{K.~Donato}
\author[csu]{W.~Fairbank Jr.}
\author[laurentian]{J.~Farine}
\author[munich]{P.~Fierlinger} 
\author[bern]{D.~Franco} 
\author[bern]{G.~Giroux}
\author[bern]{R.~Gornea}
\author[carleton]{K.~Graham}
\author[stanford]{G.~Gratta} 
\author[carleton]{C.~Green} 
\author[carleton]{C.~Hagemann}
\author[csu]{K.~Hall}
\author[laurentian]{D.~Hallman}
\author[carleton]{C.~Hargrove}
\author[slac]{S.~Herrin}
\author[ua]{M.~Hughes} 
\author[slac]{J.~Hodgson} 
\author[bern]{F.~Juget}
\author[itep]{A.~Karelin}
\author[iu]{L.J.~Kaufman}
\author[itep]{A.~Kuchenkov}
\author[umass]{K.~Kumar}
\author[seoul]{D.S.~Leonard}
\author[bern]{G.~Lutter}
\author[slac]{D.~Mackay}
\author[ua]{R.~MacLellan}
\author[munich]{M.~Marino} 
\author[csu]{B.~Mong}
\author[stanford]{M.~Montero~D\'{i}ez}
\author[umass]{P.~Morgan} 
\author[stanford]{A.R.~M\"{u}ller} 
\author[stanford]{R.~Neilson} 
\author[slac]{A.~Odian}
\author[stanford]{K.~O'Sullivan}
\author[ua]{A.~Piepke}
\author[umass]{A.~Pocar}
\author[slac]{C.Y.~Prescott}
\author[ua]{K.~Pushkin}
\author[stanford]{A.~Rivas}
\author[carleton]{E.~Rollin}
\author[slac]{P.C.~Rowson}
\author[stanford]{A.~Sabourov}
\author[sinclair]{D.~Sinclair}
\author[slac]{K.~Skarpaas} 
\author[itep]{V.~Stekhanov}
\author[triumf]{V.~Strickland\footnote{at Carleton University.}}
\author[slac]{M.~Swift} 
\author[stanford]{K.~Twelker}
\author[bern]{J.-L.~Vuilleumier}
\author[bern]{J.-M.~Vuilleumier}
\author[bern]{M.~Weber}
\author[laurentian]{U.~Wichoski}
\author[slac]{J.~Wodin}
\author[umass]{J.D.~Wright}
\author[slac]{L.~Yang}
\address[umd]{Physics Department, University of Maryland, College Park MD, USA}
\address[laurentian]{Physics Department, Laurentian University, Sudbury ON, Canada}
\address[bern]{LHEP, Physikalisches Institut, University of Bern, Bern, Switzerland} 
\address[stanford]{Physics Department, Stanford University, Stanford CA, USA}
\address[itep]{Institute for Theoretical and Experimental Physics, Moscow, Russia}
\address[csu]{Physics Department, Colorado State University, Fort Collins CO, USA}
\address[slac]{SLAC National Accelerator Laboratory, Menlo Park CA, USA} 
\address[umass]{Physics Department, University of Massachusetts, Amherst MA, USA}
\address[carleton]{Physics Department, Carleton University, Ottawa, Canada}
\address[munich]{Technical University Munich, Munich, Germany}
\address[iu]{Physics Department and CEEM, Indiana University, Bloomington IN, USA}
\address[seoul]{Physics Department, University of Seoul, Seoul, Korea}
\address[ua]{Dept. of Physics and Astronomy,University of Alabama, Tuscaloosa AL,USA}
\address[sinclair]{Physics Department, Carleton University, Ottawa and TRIUMF, Vancouver, Canada}
\address[triumf]{TRIUMF, Vancouver, Canada}


\begin{abstract}

We describe purity measurements of the natural and enriched xenon stockpiles used by the EXO-200 double beta decay experiment based on a mass spectrometry technique. The sensitivity of the spectrometer is enhanced by several orders of magnitude by the presence of a liquid nitrogen cold trap, and many impurity species of interest can be detected at the level of one part-per-billion or better. We have used the technique to screen the EXO-200 xenon before, during, and after its use in our detector, and these measurements have proven useful. This is the first application of the cold trap mass spectrometry technique to an operating physics experiment.

\end{abstract}

\end{frontmatter}

\section{Introduction}

The EXO collaboration is constructing and operating a series of experiments to search for the neutrinoless double beta decay of $^{136}$Xe\cite{danilov}. The first such experiment, known as EXO-200, is currently collecting data at the WIPP facility near Carlsbad, New Mexico\cite{ichep2010}. EXO-200 is sensitive to a neutrinoless double beta decay half life of $6.4 \times 10^{25}$ years, equivalent to a Majorana neutrino mass of order $\sim$100 meV.

The EXO-200 detector is a liquid xenon TPC. Ionization in the liquid xenon is collected by two anode grids, and xenon scintillation (at 178 nm) is observed by large area avalanche photodiodes (LAAPDs)\cite{russell}. To ensure that radioactive backgrounds in the detector materials do not obscure the double beta decay signal, a comprehensive materials screening program was employed during the construction of the experiment\cite{doug}. This program measured and certified the radiopurity of all passive detector materials, including the TPC instrumentation, the cabling, the xenon vessel, and the shielding materials.

In this article we describe a complementary program to study the purity of the xenon source isotope itself. This campaign has been carried out before and after the natural xenon commissioning run in December 2010 - January 2011, and more recently during the first detector operations with enriched xenon. These measurements have allowed us to verify that both xenon stockpiles are suitable for their intended purposes in EXO-200, to monitor the performance of the xenon gas purifiers, and to screen for radioactive backgrounds from $^{85}$Kr and $^{39}$Ar beta decay.
 
 

The measurements described in this article are based on a cold trap mass spectrometry technique which was developed to study O$_2$, N$_2$, and CH$_4$ impurities in xenon\cite{coldtrap}. More recently the method has been extended to krypton impurities as well\cite{krypton}. The method was originally demonstrated in a bench-test environment using large quantities of xenon gas. In this paper we describe how we have adapted the technique and applied it to a working neutrino physics experiment, and we report our results.

Mass spectrometry has several attractive features as an analysis technique. It can detect both electronegative and  non-electronegative impurity species, including O$_2$ and some of the problematic noble gases which contain radioactive isotopes. It allows each impurity species to be identified and counted individually. It also allows the purity of the xenon to be screened, and possibly corrected, prior to detector operations, which is similar to how all other EXO-200 detector materials are treated.  On the other hand, this method does not directly measure the electron lifetime of the liquid xenon in the detector vessel, nor can it be used to continuously monitor the xenon purity at all times like EXO's Gas Purity Monitors\cite{GPMs}. It is also not sensitive to short-lived radioactive species like $^{222}$Rn. Despite these drawbacks, we have found the method to be useful for many purposes. 

\section{Purity requirements}
\label{sec:goals}

Both the natural xenon and enriched xenon stockpiles should be relatively free of electronegative impurities for successful TPC operations. For example, the O$_2$ concentration should be less than $\sim$1 ppb in order to collect ionization over the 20 cm drift length of the detector\cite{bakale}. Other eletronegative impurity species may have larger or smaller attachment coefficients. For example, the attachment coefficient of N$_2$, measured in liquid Ar, is about a factor of 1000 smaller than that of O$_2$\cite{biller}. 

Krypton and argon impurities are problematic because they include the beta emitters $^{85}$Kr and $^{39}$Ar. For the natural xenon stockpile, which is intended for detector commissioning only, these radioactive backgrounds are not critical, provided that the event rate does not exceed the capabilities of the DAQ system.  For the enriched xenon, however, the constraints  are less trivial. With Q values of 687 keV and 565 keV respectively, these decays could obscure a significant portion of the two neutrino double beta decay spectrum, which has a broad maximum at 800 keV in $^{136}$Xe. 

If we require that the total decay rate of each of these two backgrounds does not exceed the two neutrino decay rate of $^{136}$Xe, then using the standard isotopic abundances ($\sim 2 \times 10^{-11}$  $^{85}$Kr/$^{nat}$Kr\cite{kr85_review} and $\sim 8 \times 10^{-16}$ $^{39}$Ar/$^{nat}$Ar\cite{ar39}), the published limit\footnote{These purity requirements were formulated prior to the measurement of the $^{136}$Xe two-neutrino decay rate by EXO-200 reported in Ref. \cite{2nbb-prl}.} for the two neutrino decay rate ($1.0 \times 10^{22}$ years\cite{dama}), and the $^{136}$Xe enrichment factor of 80\%, we find that the concentration of natural krypton and argon in the enriched xenon should be less than 27 parts-per-trillion g/g and 7.9 parts-per-million g/g, respectively. Clearly the krypton goal is much more stringent than the argon goal, so krypton contamination is the more serious concern. It should also be noted, however, that these goals are quite conservative, since the two-neutrino spectrum of $^{136}$Xe extends up to 2457 keV, well beyond the endpoint energies of the these beta decay backgrounds.


Note that the mass spectrometry technique does not detect $^{85}$Kr and $^{39}$Ar directly, but instead detects the stable and abundant isotopes $^{84}$Kr and $^{40}$Ar. The concentrations of the radioactive components can then be inferred from the isotopic abundances, when these are known. Radon is another radioactive noble gas which is a serious concern. However, radon cannot be detected by our method, hence we do not consider it further in this article.

\section{Methodology}

Our analysis technique is similar to the method described in Refs. \cite{coldtrap} and \cite{krypton}. We use a residual gas analyzer (RGA) mass spectrometer to measure the impurity content of xenon gas. The RGA operates under a high vacuum ( $10^{-5}$ torr or less), so only modest amounts of xenon gas can be admitted into the device at one time. This is accomplished with a precise vacuum leak valve. The RGA measures the partial pressures of the relevant atomic masses, and these partial pressures are proportional to both the concentration of the various gas species and to their flow rate through the leak valve. By opening the leak valve further and further, we can increase the flow rate to an arbitrarily high value, resulting in higher and higher RGA partial pressures for all species. In principle this allows very small concentrations of impurities to be detected above background levels. In practice, however, the use of very high flow rates will cause the RGA to saturate due to the high partial pressure of the bulk xenon gas. Once the pressure of the xenon rises above $10^{-5}$ torr, the RGA will be unable to operate, and this limits the impurity sensitivity of the RGA to about one part-per-million. 

Since we desire to detect impurities at the part-per-billion level or better, we must prevent RGA saturation by removing the bulk xenon from the sampled gas.  This is accomplished by placing a liquid nitrogen cold trap between the leak valve and the RGA. At the output of the cold trap, the xenon partial pressure is held fixed at the vapor pressure of xenon ice at 77 K ($1.8 \times 10^{-3}$ torr), independent of the leak valve setting and the flow rate. The xenon partial pressure is further reduced below $10^{-5}$ torr at the RGA by including a high impedance element in the vacuum plumbing after the cold trap. With the xenon pressure now held fixed, we can increase the flow rate, and therefore the partial pressures of many impurity species, by several orders of magnitude. In laboratory bench tests this method has achieved sensitivity to 0.12 ppb of O$_2$ and 0.3 ppt of Kr (g/g). Impurities which do not pass through the cold trap at liquid nitrogen temperature, such as H$_2$O or heavy hydrocarbons, cannot be detected in this way, but in some cases they may be detectable as an excess background level in the the cold trap plumbing after it returns to room temperature.

\begin{figure}[t!]\centering
\includegraphics[width=120mm]{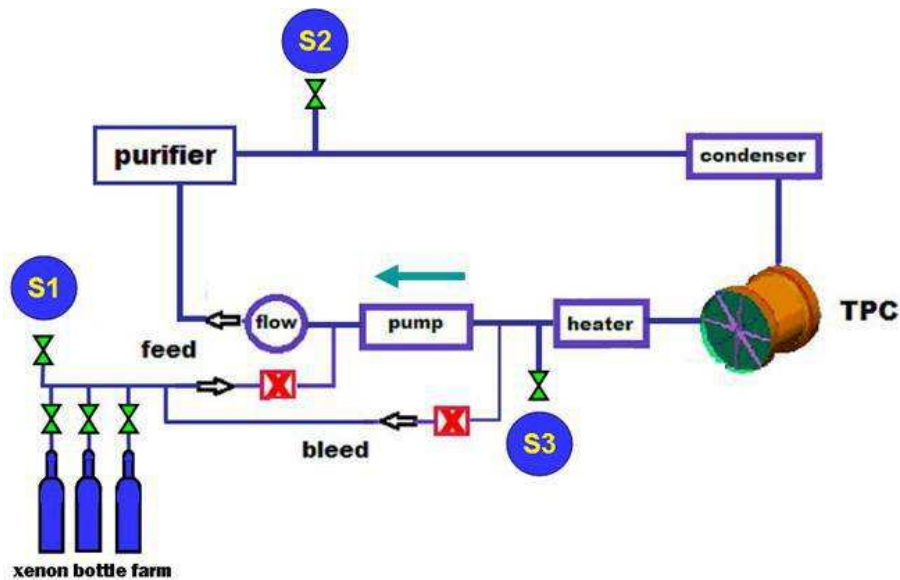}
\caption{Schematic diagram of the EXO-200 xenon handling system. We collect gas samples from S1 (xenon bottle farm), S2 (after purifier), and S3 (TPC gas return line).}
\label{fig:exo-schematic}
\end{figure}

For reasons of simplicity, we constructed the xenon gas analysis system as a stand-alone device separate from the EXO-200 xenon handling system, which is shown schematically in Figure \ref{fig:exo-schematic}. Xenon gas is collected from the various locations of interest, typically the xenon gas cylinders (S1), the output of the xenon gas purifier (S2), or the output of the TPC (S3). The gas sampling is done by attaching a 0.5 L stainless steel sample bottles to a sampling port through a valve. In the case where we sample a high pressure source like a gas cylinder, the sampling port is the output of a pressure regulator. The sample bottle is evacuated to less than $10^{-7}$ torr with a turbo pump, and then filled with xenon gas at room temperature. The pressure of the xenon gas in the sample bottle is set either by the regulator (for a high pressure source) or by the pressure of the EXO-200 xenon gas system. In either case the sample pressure is typically between 0.5 and 1.5 atmospheres. The sample bottle is then disconnected and installed on the analysis system.

\begin{figure}[t!]\centering
\includegraphics[width=120mm]{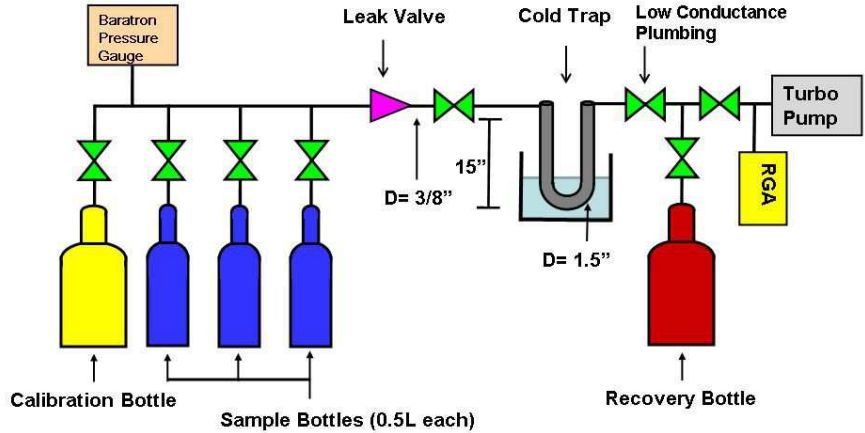}
\caption{Schematic diagram of the mass spectrometry system. Three half liter sample bottles feed xenon gas into the liquid nitrogen cold trap through a leak valve. The RGA analyzes the species which pass through the trap unimpeded. The calibration bottle contains xenon with known levels of O$_2$, Ar, and Kr impurities.}
\label{fig:analysis-schematic}
\end{figure}

A schematic of the analysis system is shown in Figure \ref{fig:analysis-schematic}. It consists of a bottle manifold, a capacitive manometer, a vacuum leak valve, a U-shaped liquid nitrogen cold trap, a low-conductance plumbing element, an MKS eVision RGA, a cold cathode vacuum gauge, and a turbo pump. The capacitive manometer (MKS Baratron model 627B)  measures the pressure in the sample bottle before and during the impurity measurement. The leak valve (Kurt J. Lesker part number VZLVM940R) admits the sampled gas to the cold trap at much reduced pressure and allows for control of the flow rate into the trap. The cold trap consists of a U-shaped segment of 3.8 cm OD stainless steel tube with a height of 38 cm designed to be immersed in a liquid nitrogen bath. The low conductance element is a $\sim 20$ cm segment of 0.6 cm and 0.95 cm OD plumbing with several right angle bends. A stainless steel bottle is attached to the output of the cold trap to allow the sampled xenon to be recovered and stored after the conclusion of the impurity measurement. The cold cathode gauge allows the total pressure at the RGA to be monitored and recorded. We have found it necessary to limit this pressure to less than $1.0 \times 10^{-5}$ torr to prevent RGA saturation effects. 

The analysis system includes a port with a pressure regulator to allow one or more sample bottles to be filled with calibration xenon. We prepared the calibration xenon gas with known concentrations of O$_2$, Ar, and Kr, by mixing known quantities of these impurities with a known amount of xenon. Except where otherwise noted, all of our impurity measurements are quantified by comparing to such calibration xenon.

\begin{figure}[t!]\centering
\includegraphics[width=120mm]{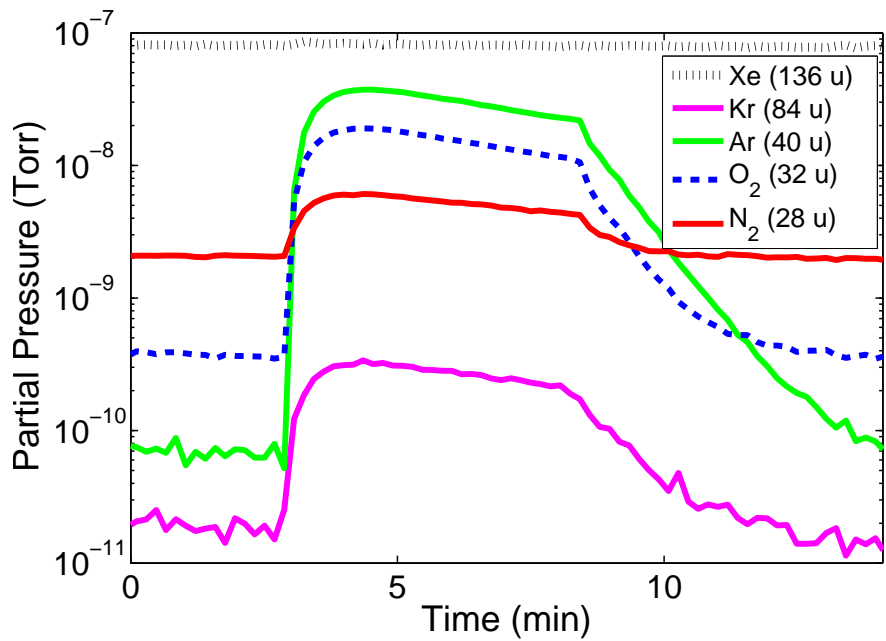}
\caption{RGA data from a typical measurement of the calibration xenon. Xenon ice was established in the trap prior to t = 0 by opening the leak valve to 1.2 turns. At t = 3 minutes, the leak valve is opened to 1.56 turns, and O$_2$, Ar, Kr, and N$_2$ are all clearly observed. The leak valve is closed at t = 8 minutes. The peak partial pressure of each species is taken as a figure of merit for impurity concentration, after accounting for the flow rate dependence.}
\label{fig:p-vs-t-example}
\end{figure}

Once a xenon gas sample has been collected and attached to the analysis system, we analyze the sample for impurities as follows. The RGA is prepared to measure and record the partial pressures as a function of time of the relevant species, as shown in Figure \ref{fig:p-vs-t-example}. We typically monitor, in atomic mass units, 28 (N$_{2}$), 32 (O$_{2}$), 18 (H$_{2}$O), 136 (Xe), 2 (H$_{2}$), 40 (Ar), and 84 (Kr). We cool the cold trap with liquid nitrogen while it is still being pumped to ultra-high vacuum with the turbo pump, and then the xenon gas is admitted into the cold trap in two steps. First, the leak valve is opened to 1.2 turns, which admits xenon into the cold trap at a very small flow rate, below $10^{-4}$ standard liters per minute (SLPM). Xenon ice forms in the cold trap and the analysis system is purged of residual trace impurities by the flowing gas. We wait about five or ten minutes for the backgrounds to stabilize before proceeding. Second, we open the leak valve to a larger flow rate, usually 1.56 turns corresponding to an flow rate of $\sim 0.1$ SLPM.  The higher flow rate used in this step allows impurities to be observed with high sensitivity, typically one part-per-billion or better.

The xenon gas in the sample bottle is slowly depleted as it flows into the cold trap. This causes the sample bottle pressure, and the flow rate through the leak valve, to decrease in time while the measurement takes place (see Figure \ref{fig:p-vs-t-example}). After a measurement period of about five minutes, the leak valve is closed, the cold trap is allowed to warm to room temperature, and the xenon in the cold trap and the sample bottle is collected in the recovery bottle with liquid nitrogen. It is sometimes useful to take a one final RGA mass spectrum of the cold trap at room temperature to look for impurity species such as H$_2$O, solvents, or heavy hydrocarbons which would be trapped in the cold trap at liquid nitrogen temperature. 

Note that to achieve the best sensitivity, the highest possible flow rate should be used for each measurement, limited only by RGA saturation effects. Since the xenon pressure is held fixed by the cold trap, RGA saturation can only be caused by impurity species which pass through the cold trap, such as O$_2$, N$_2$, and Ar. If the xenon gas sample is largely free of these impurities, the leak valve can be opened far beyond 1.56 turns, which results in higher flow and even greater sensitivity. Conversely, if the xenon purity is very poor, lower flow rates must be used to prevent saturation.

\section{Data analysis and calibration}

Previous experiments have shown that the RGA partial pressures of many impurity species of interest are proportional to both their concentration in the xenon gas and to the flow rate through the leak valve\cite{coldtrap}. The partial pressures divided by the flow rate therefore gives a figure-of-merit which is proportional to the impurity concentration, and this figure-of-merit can be calibrated in terms of absolute concentration using xenon gas of known purity. 

\begin{figure}[t!]\centering
\includegraphics[width=120mm]{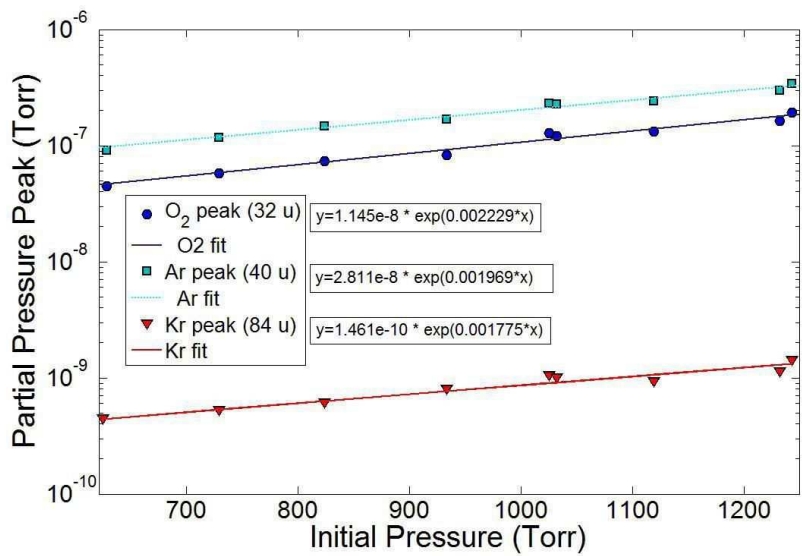}
\caption{The peak partial pressure of O$_2$, Ar, and Kr, as a function of initial sample bottle pressure for the standard leak valve position using calibration xenon. The differences in the y-axis scales for the three species reflect their concentrations in the calibration xenon. The data are described well by the indicated exponential fits.}
\label{fig:peak-pressure-calibration}
\end{figure}

The measurements described in this article differ from that of Ref. \cite{coldtrap} in that only a modest quantity of xenon gas is available for analysis in each sample, typically less than one standard liter. Therefore the sample pressure noticeably decreases during the measurement process, and so the flow rate through the leak valve changes as well. We account for the changing flow rate as follows. For a fixed leak valve setting, the flow rate as a function of time is determined by the initial xenon gas pressure in the sample bottle. Therefore the maximal values, or peak values, of the impurity partial pressures, which are realized in the first one to two minutes of the RGA measurement, also depend on the initial sample bottle pressure. As shown in Figure \ref{fig:peak-pressure-calibration}, we have measured this dependence for O$_2$, Ar, and Kr, at our nominal leak valve setting of 1.56 turns, using xenon gas from our calibration cylinder. The data is well described by a simple exponential fit, and we use this fit to derive an expression for a sample bottle pressure correction factor (CF):

\begin{equation}
 CF(\Delta p) = e^{- \alpha \Delta p}
\end{equation}

\noindent
where $\Delta p$ is the initial sample pressure minus 1000 torr, and $\alpha$ is $2.22 \times 10^{-3}$, $1.97 \times 10^{-3}$, and $1.78 \times 10^{-3}$ for O$_2$, Ar, and Kr respectively (all in units of torr$^{-1}$). This correction factor accounts for the flow rate dependence of the peak partial pressure for each species.

The O$_2$, Ar, and Kr concentration of a xenon gas sample is determined by comparing the corrected peak partial pressure with the same value from a sample bottle filled with the calibration xenon at 1000 torr initial pressure:

\begin{equation}
	\rho(\rm{sample}) = \frac{CF \times PP(\rm{sample})}{PP(\rm{calib @ 1000 torr})} \times \rho(\rm{calib})
\end{equation}

\noindent
where $\rho$ is the concentration of the impurity and  $PP$ is the peak partial pressure. The calibration xenon has been prepared with 1.0 part-per-million (ppm) O$_2$, 1.6 ppm Ar, and 42 part-per-billion (ppb) Kr, in units of g/g. 

\begin{table}[t!]
\small
\begin{centering}
\begin{tabular}{|c|c|c|}
\hline
Impurity species & AMUs &  response factor relative to O$_2$  \\ 
\hline
He&4&5.0\\ \hline
$\rm CH_4$&15&3.0\\ \hline
$\rm N_2$&28&1.6\\ \hline
Ar&40&1.1\\ \hline
Kr&84&0.25\\ \hline
\end{tabular}
\caption{Analysis system response factors for several species, relative to O$_2$, in units of g/g.}
\label{tab:spec_cal}
\end{centering}
\end{table}

Our calibration xenon does not contain CH$_4$ or helium, and the amount of N$_2$ is not independently known, so for these species we make two adjustments. First, we correct for the flow rate dependence at our nominal leak valve setting by assuming the input-pressure correction curve for O$_2$. We take a systematic error of 10\% on this correction. Secondly, to account for the species-dependent response of the analysis system, including factors such as the probability that each impurity survives the cold trap and its ionization potential at the RGA filament, we apply a relative species response factor as listed in Table \ref{tab:spec_cal}. These species response factors were empirically determined by comparing each species to O$_2$ in separate bench test measurements at constant input pressure, and are valid to within 30\%.

\begin{figure}[t!]\centering
\includegraphics[width=120mm]{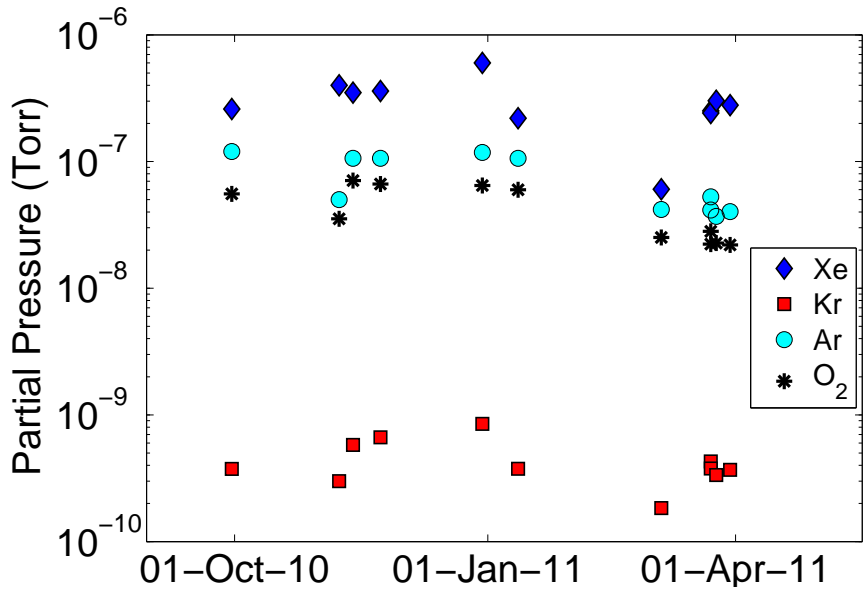}
\caption{Peak partial pressure measurements of the calibration xenon from 9/30/2010 to 3/30/2011, corrected to 1000 torr input pressure equivalent. Kr, Ar, and O$_2$ correspond to a fixed concentration of 42.4, 1200 and 1000 $ \times 10^{-9}$  (g/g) respectively. The variations are believed to be dominated by the drift of the RGA gain.}
\label{fig:RGA-calibration-drift}
\end{figure}

To monitor for daily variations in the system response, including changes in the RGA gain, we periodically calibrate the analysis system using our cylinder of calibration xenon gas. Figure \ref{fig:RGA-calibration-drift} shows the variations in the system response to various impurities for our calibration xenon over a period of six months. To account for these variations, we use the average calibration for each species before and after the date of the gas sample analysis, and we take the the deviation from the average to be a systematic uncertainty. For most of our measurements, this is the dominating uncertainty in the impurity concentration measurement.

\section{Results}

\subsection{Purity measurements from the EXO-200 natural xenon run}

EXO-200 carried out a commissioning run in December 2010  through January 2011 using natural (unenriched) xenon. This xenon was supplied by the gas vendor in six compressed gas cylinders, five of which contained 35 kg of xenon, and one of which contained 50 kg. We analyzed each of the cylinders with the cold trap mass spectrometry technique prior to filling the EXO-200 TPC. After the commissioning run was completed, the xenon was recovered into the gas cylinders using two xenon gas compressors. One final sample of the mixed xenon was taken and analyzed after the recovery was completed. Results from all of these measurements are shown in Table \ref{tab:Natural_Xe}.

\begin{table}[t!]
\begin{centering}
\begin{tabular}{|c|c|c|c|c|c|c|}
\hline
Label & Kr & Ar & $\rm N_2$ & $\rm O_2$ & He \\
 &(ppb)&(ppb)&(ppb)&(ppb) & (ppb)\\ 
\hline
E4 (35kg)    &$60.1\pm 6.7$    &$1850\pm 850$    &$32200\pm 12500$    &$14300\pm 3190$&$840\pm 270$\\ 
\hline
E6 (35kg)    &$85.4\pm 9.5$&$27.4\pm 12.6$    &$1170\pm 370$    &$<$6.5    &$588\pm 305$\\ 
\hline
E7 (35kg)    &$25.1\pm 2.8$    &$30.3\pm 13.9    $&$809\pm 256$    &$<$1.8    &$581\pm 301$\\ 
\hline
E2 (35kg)    &$32.7\pm 3.6$&$131\pm 60$    &$2490\pm 790$    &$616\pm 138$&$444\pm 230$\\ 
\hline
E8 (50kg)    &$20.7\pm 6.6$    &$2.2\pm 0.9$    &$46.9\pm 18.9$    &$<$1.9    &$8.5\pm 4.1$\\ 
\hline
E5 (35kg)    &$36.4\pm 4.4$    &$2.6\pm 0.6$    &$12.2\pm 2.7$    &$<$0.8    &$<$0.1\\ 
\hline
\hline
Stockpile avg.  &$42.6\pm 5.7$    &$319\pm 146$    &$5720\pm 2180$&$2320\pm 520$&$384\pm 173$\\
\hline
\hline
After recovery    &$42.9\pm 16.6$    &$256\pm 53$    &$108\pm 25$    &$<$0.3    &$257\pm 83$\\ 
\hline
\end{tabular}
\caption{Results of Kr, Ar, N$_2$, O$_2$, and helium concentration measurements from the EXO-200 natural xenon gas cylinders. The six gas cylinders labeled E2 through E8 were individually sampled prior to filling the detector. E2 and E4 show evidence for air contamination. See Section \ref{sec:air} for details. 'Stockpile avg.' refers to the mass-weighed average purity of all six cylinders. 'After recovery' refers to a sample extracted from the gas cylinders after the xenon was recovered from the EXO-200 TPC vessel with xenon gas compressors. The indicated uncertainties includes all systematic errors and is dominated by RGA gain drift.  ppb refers to $10^{-9}$ g/g.}
\label{tab:Natural_Xe}
\end{centering}
\end{table}

The EXO-200 natural xenon was also sampled from the TPC xenon gas return line (sample location S3 in Figure \ref{fig:exo-schematic}) while the TPC was filled with liquid xenon, and also during the xenon recovery process at the conclusion of the natural xenon run. The noble gas impurity measurements from these samples are listed in Table \ref{tab:solubility}.

\begin{table}[t!]
\begin{centering}
\small
\begin{tabular}{|c|c|c|c|}
\hline
& Kr &Ar &He\\
& (ppb)& (ppb)& (ppb)\\ \hline
stockpile avg., before liquefaction &$42.6\pm 5.7$&$319\pm 146$&$384\pm 173$ \\ \hline
port S3, detector filled with liquid& $4.4\pm 1.7$& $1.0\pm 0.2$& $<$0.04 \\ \hline
port S3, during liquid recovery& $21.8\pm 8.4$& $26.2\pm 5.4$&$<$0.06 \\ \hline
port S3, residual gas after recovery& $118\pm 46$& $1890\pm 390$& $3060\pm 980$ \\ \hline
port S1, after recovery & $42.9 \pm 16.6$ & $256 \pm 53$ & $257 \pm 83$ \\ \hline
 
\end{tabular}
\caption{Concentrations of Kr, Ar, and He, in the EXO-200 xenon handling system during and after the natural xenon commissioning run. The ``stockpile avg.'' and ``after recovery'' results are taken from Table \ref{tab:Natural_Xe}). When liquid xenon is present in the TPC, the xenon vapor above the liquid shows a deficit of these noble gas impurities.  After recovery there is an overabundance of impurities left in the residual gas in the detector.}
\label{tab:solubility}
\end{centering}
\end{table}

\subsection{Purity measurements of the EXO-200 enriched xenon}

After the recovery of the natural xenon, and a thorough pump-down of the vessel to remove the krypton impurities, EXO-200 began operations with xenon gas enriched by ultra-centrifugation to 80\% in $^{136}$Xe . The enriched xenon is stored in ten compressed gas cylinders. Because of the value of the material and the risk involved with any handling of it, most of the cylinders were not sampled individually, however the EXO-200 xenon gas handling system was sampled before and during detector filling to monitor the purification and liquefaction process. The first enriched xenon cylinder, known as E-42, was used to fill the detector vessel with $\sim$1 atmosphere of gaseous xenon prior to liquefaction. This gas was sampled from port S3 to confirm that the krypton present in the natural xenon run had been successfully removed. All ten enriched bottles were then opened and liquefaction commenced. The contamination levels of samples taken at different phases of the detector filling are listed in Table \ref{tab:Enriched_Xe}.

\begin{table}[t!]
\begin{centering}
\footnotesize
\begin{tabular}{|c|c|c|c|c|c|c|c|c|}
\hline
Sample    & Port    & Comment    &$\rm O_2$    &$\rm N_2$    &Ar    &He    &Kr    & CH$_4$    \\
&    &      & (ppb) & (ppb) & (ppb) & (ppb) & (ppb) & (ppb) \\ 
\hline    
Enr-A    &S3    & Purified TPC gas,          &$<$0.7    &$<$4.5    &128$\pm$61    &*70$\pm$36    &$<$0.4    &$<$1.9    \\ 
 & & before liquefaction. & & & & & & \\
\hline
Enr-H$^\dagger$    &S1    & Stockpile mixture,        &$<$0.4    &329$\pm$81    &8.9$\pm$2.0    &*42$\pm$14    &0.0235$\pm$0.004    &24.9$\pm$6.1    \\ 
 & & before purification. & & & & & & \\
\hline
Enr-F    &S2    &   Start of liquefaction,      &$<$4    &$<$28    &9.5$\pm$2.0    &1.5$\pm$0.5    &0.0517$\pm$0.007    &1.3$\pm$0.3    \\
 & & after purifier. & & & & & & \\
\hline
Enr-G    &S2    & During liquefaction,        &$<$3    &$<$27    &7.6$\pm$1.8    &1.3$\pm$0.4    &0.038$\pm$0.007    &1.7$\pm$0.4    \\ 
 & & after purifier. & & & & & & \\
\hline
Enr-I & S1 & Dedicated Kr search,  &  & & & & 0.0274 $\pm$ 0.004 & \\ 
 & & repeat of Enr-H. & & & & & & \\
\hline
\end{tabular}
\caption{Results of selected purity measurements of the EXO-200 enriched xenon. ppb refers to $10^{-9}$ g/g. The samples are listed in chronological order, although the labeling is out of order. *The helium concentration in these samples may be erroneously elevated due to a small amount of helium remaining in the plumbing following a leak test procedure. $^\dagger$ CF$_4$ was also seen in this sample.}
\label{tab:Enriched_Xe}
\end{centering}
\end{table}

\section{Discussion}

\subsection{Electronegative impurities in the natural xenon stockpile} 
\label{sec:air}

As shown in Table \ref{tab:Natural_Xe}, four of the EXO-200 natural xenon gas cylinders supplied by the vendor contained less than 1 part-per-million of N$_2$ and less than one part-per-billion of O$_2$. This purity level is remarkable, because it is comparable to that required for successful TPC operations, even without further purification. Cylinders E2 and E4, however, show modest amounts of N$_2$, O$_2$, and Ar, consistent with air contamination. Furthermore, analysis of two other EXO-200 natural xenon gas cylinders, not listed in Table \ref{tab:Natural_Xe}, had much larger amounts of air, estimated to be hundreds of parts-per-million. The air contamination was likely introduced during an EXO cryogenics commissioning run in 2009, when one of the xenon recovery compressors had a leaky valve.

From the concentration measurements shown in Table \ref{tab:Natural_Xe}, we calculate that the total mass of N$_2$ and O$_2$ in the unpurified natural xenon stockpile was 1.3 grams and 0.5 grams, respectively. These modest quantities are readily removed by the zirconium getters without becoming exhausted\cite{purifier}. Therefore these six cylinders were certified for use in EXO-200. The two cylinders with more significant air contamination, not shown in Table \ref{tab:Natural_Xe}, would have easily overwhelmed the purifiers and exposed the TPC to large quantities oxygen, and so they were returned to the gas vendor for reprocessing.

\subsection{Natural xenon purity after recovery}

After the conclusion of the natural xenon commissioning run, the xenon was recovered into the six storage cylinders and sampled again. As shown in Table \ref{tab:Natural_Xe}, the concentrations of the noble gas impurities remained the same before and after the run, as expected, since they are not removed by the zirconium getter. The concentration of O$_2$ and N$_2$ was significantly reduced after the recovery, having been removed by the zirconium getter during detector filling and re-circulation. However, some detectable N$_2$ remains after the recovery, most likely from outgassing from the acrylic and teflon TPC components. No detectable O$_2$ is found after recovery, which validates the integrity of the recovery plumbing, including the xenon compressors. 

\subsection{Krypton and argon content of the natural xenon}

Krypton and argon impurities were detected in the natural xenon before, during, and after the commissioning run. As shown in Table \ref{tab:solubility}, the average krypton and argon concentrations in the natural xenon stockpile are $\rm 42.6 \pm 5.7$ ppb (g/g) and $202 \pm 20$ ppb (g/g), respectively. The  $^{85}$Kr beta decay was, in fact, observed in the TPC data during the natural xenon commissioning run. The decay rate observed in the TPC is consistent with the mass spectrometry result, assuming the standard $^{85}$Kr isotopic abundance of $\sim 10^{-11}$. Note that this krypton level is acceptable, since the natural xenon is only used for detector commissioning.

It is also apparent from Table \ref{tab:solubility} that in samples of xenon gas taken above the liquid, the noble gas impurity concentrations appear suppressed below their average values for the homogeneous gas phase. We find that the krypton concentration is reduced by a factor of 10, argon by a factor of 200, and helium by more than a factor of 6000. This could indicate that the solubility of these impurities in liquid xenon is rather large.

Conversely, after the xenon was recovered into the storage cylinders at the conclusion of the natural xenon run, the residual gas left behind in the TPC vessel was sampled and shown to be enriched in the noble gas impurities (see Table \ref{tab:solubility}). This over-abundance is presumably due to the distillation effect of the recovery process. This krypton measurement, in particular, confirmed that it would be necessary to evacuate the TPC vessel before filling with enriched xenon.

\subsection{Electronegative content of the enriched xenon}

As shown in Table \ref{tab:Enriched_Xe}, no O$_2$ was observed in the un-purified enriched xenon, with a limit of $<$ 0.4 parts-per-billion g/g being set, but N$_2$ was observed at a level of $329 \pm 81$ parts-per-billion g/g, for a total N$_2$ burden of 0.06 grams. This amount was easily removed by the purifiers, as confirmed by subsequent measurements at port S2. Methane was also observed in the unpurified enriched xenon, and a small amount remained at the purifier output, consistent with the expected purification efficiency for that species\cite{purifier}. Measurements performed by our Gas Purity Monitors, on the other hand, did find positive evidence for electronegative species in the purified enriched xenon, in contrast to the cold trap measurements discussed here. Those GPM measurements and related studies will be described in a future publication.

\subsection{Krypton and argon content of the enriched xenon}

The centrifugation process is expected to efficiently remove krypton from the enriched xenon stockpile, and indeed a sample of purified enriched xenon gas collected from the TPC prior to liquefaction placed a limit of 0.4 ppb on the krypton concentration (see sample Enr-A in Table \ref{tab:Enriched_Xe}). Argon and helium were seen in this sample, which is surprising since these species should also be removed by the centrifuges.

To improve our sensitivity to much smaller concentrations of krypton, we performed additional analysis of enriched xenon samples using the largest flow rate which can be achieved with our vacuum leak valve, which is about 100 times higher than our standard flow rate. We account for the increased flow rate by measuring the peak partial pressure of krypton and dividing by the average flow rate observed in the first 30 seconds of each dataset. As usual, the response of the cold trap and RGA to krypton was calibrated for this leak valve position using the calibration xenon. 

\begin{figure}[t!]\centering
\includegraphics[width=70mm]{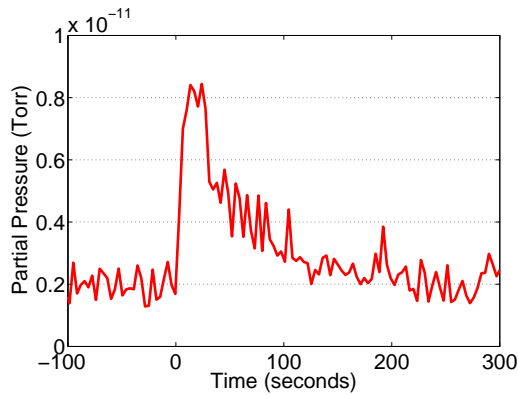}
\caption{Detection of krypton in the enriched xenon at high flow rate (sample Enr-H). The signal drops rapidly after peaking due to the rapidly decreasing input pressure which is realized with the large leak rate. Comparing to the calibration xenon gives a krypton concentration of 23.5$\pm$4 parts-per-trillion (g/g). }
\label{fig:EXe_Kr}
\end{figure}

As listed in Table \ref{tab:Enriched_Xe}, and shown in Figure \ref{fig:EXe_Kr}, gas sample Enr-H found strong evidence for the presence of krypton in the enriched xenon at a concentration of  23.5 $\pm$ 4 parts-per-trillion (g/g). A second measurement (Enr-I) found a krypton level of 27.4$\pm$4 parts-per-trillion (g/g). Averaging the two values we find the value to the krypton concentration in the enriched xenon to be 25.5$\pm$3 parts-per-trillion (g/g). 

Taken at face value, these measurements satisfy our goals for krypton and argon impurities in the enriched xenon. However, an important caveat should be noted. Since our mass spectrometry technique measures the stable isotopes of krypton and argon ($^{84}$Kr and $^{40}$Ar), the radioactive components must be inferred from knowledge of the isotopic abundances. In the case of the EXO-200 enriched xenon, the standard isotopic abundances may be in error for two reasons. First, the radioactive components decrease as the xenon ages, and the date of bottling is not known very well. Secondly, the centrifugation process will increase the fraction of the heavy isotopes relative to the light ones. In the case of argon, both effects tend to reduce the $^{39}$Ar abundance (satisfying our goal even more strongly), but for $^{85}$Kr the enrichment effect works opposite to the aging effect, and the final result is not precisely known.  Measurement of the other stable isotopes of krypton indicates that the $^{86}$Kr/$^{84}$Kr ratio may be enhanced by as much as 50\% relative to its natural value, but with a large statistical uncertainty. This places a loose constraint on the krypton enrichment effect in our xenon. We conclude that the $^{85}$Kr abundance is comparable to our goal, although a precise determination is not possible.

\section{Conclusion}

We have applied the cold trap mass spectrometry technique to study the purity of the xenon stockpiles of the EXO-200 double beta decay experiment. Compared to Ref. \cite{coldtrap}, we have modified the technique to allow the analysis of relatively small samples of xenon gas, typically less than one standard liter. This is the first application of this method to a working physics experiment.

We find that the method is most useful when samples of the xenon gas are collected and analyzed before liquefaction and after recovery. Samples collected from the xenon vapor during detector operations, on the other hand, found anomalously low noble gas impurity levels, as shown in Table \ref{tab:solubility}. We infer that these species are missing from the gas phase due to solvation effects, and this places some limitation on the usefulness of mass spectrometry data collected during detector operations. 

The measurements have nevertheless provided much useful guidance as we have filled and operated EXO-200 for the first time. Our natural xenon screening campaign verified the suitability of the EXO-200 natural xenon stockpile, and it allowed us to isolate two natural xenon cylinders whose impurity content would have saturated our purifiers. We measured the krypton content of the both the natural and enriched xenon prior to filling the detector. We also found that the unpurified enriched xenon was largely free of O$_2$, but contained modest quantities of N$_2$ and methane.

\section{Acknowledgments}

EXO is supported by DoE and NSF in the United States, NSERC in Canada, SNF in Switzerland and RFBR in Russia. 

\bibliographystyle{elsarticle-num}
\bibliography{sampling}

\end{document}